# Comprehensive Digital Forensics and Risk Mitigation Strategy for Modern Enterprises

Shamnad Mohamed Shaffi

Colorado Technical University
CS631 - 2004A - 01: Digital Forensics

**Abstract:** *Enterprises today face increasing cybersecurity threats that necessitate robust digital forensics and risk mitigation strategies. This paper explores these challenges through an imaginary case study of Entity Proxy (EP), a global identity management and data analytics company handling vast customer data. Given the critical nature of its data assets, EP has established a dedicated digital forensics team to detect threats, manage vulnerabilities, and respond to security incidents. This study outlines EP's approach to cybersecurity, including proactive threat anticipation, forensic investigations, and compliance with regulations like GDPR and CCPA. Key threats such as social engineering, insider risks, phishing, and ransomware are examined, along with mitigation strategies leveraging AI and machine learning. By detailing EP's security framework, this paper highlights best practices in digital forensics, incident response, and enterprise risk management. The findings emphasize the importance of continuous monitoring, policy enforcement, and adaptive security measures to protect sensitive data and ensure business continuity in an evolving threat landscape*

## 1. Company Overview

Entity Proxy (EP) is a leading identity management company that offers data, technology, ethics, and organizations worldwide ideas to build their customer - centric business. The Company was established in 2015 and had its headquarters in Seattle, Washington state. The Company provides digital solutions that bring together customer pre - calculated attributes across multiple first - party systems and incorporate third - party demographics at individual and household and supply data to append a holistic view of an organization's customers. The Company also leverages artificial intelligence tools and machine learning to research customer journeys, customer interaction patterns, and update the identified model for accuracy and data integrity. Organizations use this data to build a customer journey platform that helps business users understand all kinds of customers and commonly asked for Customer attributes pre - calculated and maintained. The Entity proxy has 1500 full - time employees, around 100 contractors, and consultants, and supports more than 100 clients across 20 countries.

### a) Foundational capabilities and Research Areas
Entity Proxy digital solutions fall within one of three foundational capabilities. Each foundational office maintains data used for different targeted audiences.

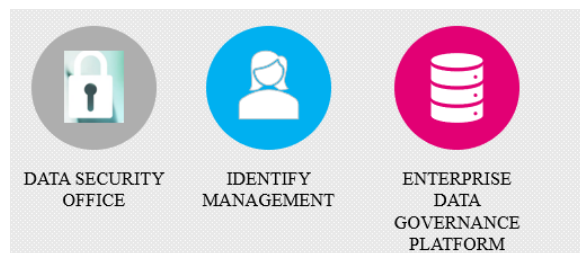

The identify management team manages and maintains a customer Profile application that provides a snapshot description of the Customer; the information can be deterministic (e. g., name, age) or calculated. Supports products such as decision engine, customer segmentation. The team also builds a Relationship Network platform that provides the link between customer entities and their nature. It also supports regulatory requirements such as CCPA and GDPR. The data security office's vision is to foster capabilities that govern and protect the sensitive data, and their strategy is to mature and deliver capabilities that provide data security through discovery, protection, and governance. The Enterprise Data Governance platform is a data management and governance platform owned by the Data Governance Office (DGO) that helps organizations to discover, understand, manage, govern, and use our enterprise data.

### b) Digital Forensics Plan Goal
Entity Proxy must have mature assessment programs that effectively discover vulnerabilities, analyze the potential risk and impact of vulnerabilities, prioritize, and manage remediation. Digital business with associated trends and drivers such as mobility, the Internet of Things (IoT), and cloud increases the proliferation of advanced threats and is prompting a shift in mindset from prevention to "detect and respond, " changing the focus from pure risk reduction to risk management. Any breach in customer data will impact the business, damaging the stakeholder relationships, and imposing hefty fines due to compliance requirements.

### c) Vulnerabilities and Root Causes
Entity proxy maintains Individual demographics data, and any data breach would result in compromising customer data. The Company has operations on multiple services, and the enterprise data is not consolidated into a centralized data warehouse. Customer sensitive information could be saved in multiple applications in clear text format, and this adds complexity and challenge in fetching the subscriber data to support the data request from customers. The Company also has contractors working on multiple projects with access to sensitive data, which is a significant security concern. The Company has data pipelines and handshakes created between the clients and needs to ensure data security in motion and rest.

#### d) Digital Forensics, Containment, and Minimizing Risks

Digital forensics is a scientific process for validating, identifying, analyzing, and reconstructing the digital and computer - based events while preserving the integrity of the original artifacts to prosecute criminal individuals or organizations that attempt to disrupt business operations (Gladyshev, 2004). Sufficient threat and vulnerability management can maximize an enterprise's security posture while minimizing the resources required to do so. To ensure success, IT security leaders must verify the effectiveness of their vulnerability management (VM) efforts and align these with business context and objectives. Modeling the impact of potential threats to evaluate their risk and enabling more targeted mitigation efforts will become a primary tool in managing the large volume of vulnerabilities that enterprises typically detect. The digital forensic team should have processes and controls in place to Anticipate threats, and their impact will be critical to prioritize and remediate vulnerabilities before they are exploited. The incident response team should effectively coordinate with internal and external stakeholders on identifying, impact assessment, remediation, Measures and measurements, and communication when a breach happens in the organization.

## 2. Threats and Vulnerabilities

Entity proxy has Individual demographics data, and any data breach would result in compromising customer data. Hackers can get access to sensitive and confidential information leverage the tools and technology techniques such as remote, local, client, or denial of service.

### 2.1 Remote

Entity proxy has its information data assets stored on both cloud and on - premises data centers. Malicious actors can use remote access tools and convince employees that they work for our support IT department. Once they have established a remote session, they can commit fraudulent actions like installing malware, credential/password scrape from browsers, etc.

#### 1) Social Engineering
Social engineering is another way that an attacker may gain information that can assist in cracking a password. (MUSE, n, d). Entity proxy has full - time employees, contractors, and external consultants working on different projects. Social Engineering is the primary method by which companies are attacked and breached. A social engineering attack is an orchestrated drive against employees of companies using a variety of techniques to steal the Company's intellectual property and assets.

#### 2) Phishing / Spear Phishing
Another potential attack is that phishing is another type of social engineering attack. Phishing goes after the most vulnerable part of any security system, the People. Phishing is only useful if the user takes the hook and clicks.

### 2.2 Local

Local attacks to the information assets exploit vulnerabilities within the data centers located at various physical locations. The major local vulnerabilities are unauthorized physical access to physical locations, internal or insider threats, and violating its non - compliance policies.

#### 1) Unauthorized Physical Access
There must be physical security systems at enterprise facilities owned, rented, or leased by the Company and apply to persons on Company property. Employees, contractors, and vendors working in or accessing company facilities have specific security - related responsibilities and shall comply with the Company's security policies, standards, procedures, and guidelines. An unauthorized individual can enter or attempt to enter facilities or work areas where they do not have a legitimate business need. Other means of unauthorized access permit piggybacking, tailgating, access badge sharing, identity misrepresentation, bypassing turnstiles, manipulating locks, or other means.

#### 2) Insider Threats
Insider threats come in three flavors. Accidents happen, and employees fall for phishing scams or mistakenly expose confidential information. Negligence can also be a threat; attempts to circumvent or ignore policies and best practices because it makes life easier can create vulnerabilities. For instance, using your unsecured device on the network can create an avenue for hackers' attacks. The most severe insider threat is a malicious employee. Often, cybercriminals will recruit insiders to help them execute an attack. Recruiters use social engineering and social media reconnaissance to blackmail an employee into becoming an agent. They might also prey on willing recruits who are motivated by money or revenge.

#### 3) Intentional or Accidental Non - compliance
Entity Proxy has Policies and standards that outline the requirements for the acceptable use of its information and data Resources. It is intended to help ensure the protection and legitimate use of Entity Proxy's business assets, including employee, Customer, and other company information. Any use to access, display, create, transmit, receive, or store material, content, or images that violate applicable laws and/or company policies. Applicable laws and policies include those governing workplace discrimination, intellectual property, trade secrets, fair trade practices, etc.

### 2.3 Client

The client - side attacks use the software or applications in the Company to exploit the resources.

Client - side or Local storage "means data stored on the Customer's browser. Hackers can use different approaches to visit a malicious website to get browser information. Watering hole attacks and cross - site scripting (XSS) or session - hijacking are a few client - side attacks.

### 1) Watering Hole Attacks

In this attach strategy, the hacker identifies the patterns which the websites an organization often visits and exploit the information by infecting with any malware. A specific IP address can be targeted for the attack, and a group of individuals can be infected using the malware.

### 2) Client & Browser Attacks (XSS Session Hijacking)

XSS is a session hijacking strategy where the web session will be exploited. The session token can be compromised in different methods like session sniffing, predictable session toke, client - side attacks, Malicious JavaScript codes, trojans, etc.

## 2.4 DoS

Denial - of - Service (DoS) attack is intended to deny access to the Company's critical information assets for intended users. Hackers can flood the network server with traffic is a common method of DoS attack.

### 1) DDoS

A distributed denial - of - service (DDoS) attack is a malicious attempt to disrupt the regular traffic of a targeted server, service, or network by overwhelming the target or its surrounding infrastructure with a flood of Internet traffic (Cloudflare, n, d).

### 2) Ransomware

This is one of the top threat vectors in the industry. For example, the breach was due to ransomware, leading to revenue loss and impacted business operations for several months. Supplementing technology solutions with an effective set of policies, procedures, and processes is an industry best practice to counteract the threat associated with ransomware attacks. CIS must define the problems and risks associated with ransomware and collaborate with the end - user and infrastructure teams to develop effective processes and technology solutions.

## 3. Computer System Incidents

Threat actors use different approaches to gain access to an enterprise network. It is critical to understand the vulnerabilities and types of breaches and events that lead to an attack. This helps identify the breach's scope and take preventive measures to contain the attack and reduce information assets' impact.

### a) The Cyber Kill Chain

The cyber kill chain is a series of steps that trace a cyberattack stage from the early reconnaissance stages to the exfiltration of data. The kill chain helps us understand and combat ransomware, security breaches, and advanced persistent attacks (Hospelhorn, 2020). The cyber kill chain has eight core stages that range from the initial stage of the attack, navigating to the network to access data exfiltration. Any standard type of accts like phishing, brute force, malware etc. will trigger a cyber kill chain activity.

| Phase | Description | Examples |
|---|---|---|
| Reconnaissance | Attackers' research gains information about the security systems in place, identifies IPS and authentication processes and selects the targets. | Harvesting email addresses, conference data |
| Intrusion | Attackers attempt to remote pairing and inject malware into the network. | Remote services |
| Exploitation | Once the attackers gain access to the network, try to exploit any vulnerabilities and threats inside the enterprise systems. | Scripting, Job scheduling |
| Privilege Escalation | In this phase, the attacker tries to get additional permissions or privileges to accounts and systems. | Manipulating Access tokens |
| Lateral Movement | The attacker attempts to access the Company's critical information assets like customer data and financial information by connecting to additional systems. | SSH hijacking |
| Obfuscation | In this stage, the attacker tries to remove the logs, timestamps, modify the security systems, etc. to hide that attack doesn't happen. | Delete files, user file deletion |
| Denial of Service | The attacker tries to disrupt a regular enterprise operation that is mainly done to distract the security breach. | Shutdown systems, stopping the services. |
| Exfiltration | At this stage, the attackers accrued all the critical enterprise data, copying the data outside the enterprise and either could sell the confidential data or upload to open sites. | Data encrypted. |

### b) Incident Response Team Responsibilities

The incident response team's goal is to coordinate and align the key resources and team members during a cybersecurity incident to minimize impact and restore operations as quickly as possible (Business, n, d). The following are the key responsibilities of an incident response team. Respond to Cyber Incidents Immediately and As Needed - Cyber incidents can and will happen outside of regular working hours. Be ready and available to respond to these incidents as they happen. Sit in on incident response calls and begin analysis.

Document Incident Details, Diagnose Issues, and Create Kill Chain - Take clear and concise notes that can be communicated up to leadership. Begin root cause analysis and craft remediation action items. Diagnose any trouble in the management flow caused by a failure of any kind and determine to remedy the causes of the symptoms, with the final product being the confirmation that the solution restores the process to an excellent working state. Draft kill chain flows/diagrams.

Project Managed Cyber Incident Remediation Efforts - Ensure remediation project is delivered on time, to budget, and to the expected quality standard. Ensure the project is effectively resourced and managed relationships with a wide range of groups/orgs. Responsible for managing the work of consultants, allocating/utilizing resources in an effective and efficient manner. Maintain a cooperative, motivated, and successful team.

Technology Trend Awareness - Must be able to look back at the industry's setbacks and successes and consider new possibilities for the future use of technology, looking for a better, faster, more practical approach that can make business more productive. Introduce innovation to business to help save time and money, giving a competitive advantage to grow and adapt the business in today's marketplace and create more efficient processes and ideas with a likelihood for the business to succeed.

*c) Threat Identification and Analysis*

The incident management team monitors alert feeds to identify suspicious activity that requires further analysis. Based on this analysis, a cyber incident may be declared and escalated per the Incident management plan. Network forensic or incident response teams use Intrusion Detection Systems (IDS) and evidence collection tools to gather the network traffic and logs to identify any network intrusion, attacks, and hacking, fraud, etc.

Key activities in this phase include -
- Collect and analyze cyber events/alerts to evaluate against incident declaration criteria▪
- Invoke the Cyber Incident Response Plan
- Perform initial analysis to support severity declaration
- Create an Incident Report and begin documentation of incident findings.
- Incident escalation to the senior leadership team
1) Local threats - Unauthorized access can be identified by enabling access logs on the applications. And need to revisit and renew the security policies and provide required security training for the employee to alert as and when a red flag is identified.
2) Client - side attacks – Client - side attacks can be identified using log analysis, file systems, static analysis, etc. Timestamps of events collected from different devices can be used to investigate and recreate the events.
3) Network traffic - Threat vectors are paths that a malicious attacker may take to compromise or bypass company defenses and infect the network. The network packets can be leveraged to identify the events related to ports and protocols.
4) Identifying DoS or DDoS - involves protection from denial - of - service attacks by continuously monitoring the network traffic and create a baseline on the typical and critical workloads. Any spike in the network traffic needs to be alerted based on the network configuration and monitoring tools that help identify the threats as early as possible.

## 4. Network and Internet Incidents

a) **IDS/IPS Analysis**
Intrusion Detection Systems (IDS) provide monitoring and visibility into the network's security posture and detect intrusions. Intrusion Prevention Systems (IPS) provide the ability to block detected intrusions and attacks. An IDS device or software application's primary objective is to monitor and alert the network or system activities for malicious activities or policy violations and produces reports to a concerned stakeholder in an organization. The monitoring framework consists of the below components.
- Controls and processes for continuous monitoring of network traffic, information assets, data in motion and rest, etc.
- System configuration error correction
- Evaluating the integrity of confidential systems and data files.
- Monitoring and identifying threat attacks based on known patterns.
- Monitoring Audit trails and alerting policy violations and vulnerabilities.

*1) Main Components of an IDS*
The following are the main components of an Intrusion Detection Systems (IDS)
- Data Collection
- Data Conversion
- Decision Engine

The data collection components are responsible for collecting data, i. e., network packets, log files, system call traces, etc. Data conversion analyzes the inputs received from the collection components. Data conversion is responsible for determining if an intrusion or attack has occurred. The decision engine enables a user to view the output from the system's system or control behavior. Data visualization tools or GUI reports may be created accessible to the managers and the leadership team.

An intrusion detection system (IDS) is classified into the below categories.
- Network intrusion detection systems (NIDS)
- Host - based intrusion detection systems (HIDS)
- Signature - based intrusion detection systems

*2) Network Intrusion Detection Systems (NIDS)*
Network - based intrusion detection systems (NIDS) are devices intelligently distributed within networks that passively inspect traffic traversing the devices on which they sit (Timothy, Shimeall, 2014).

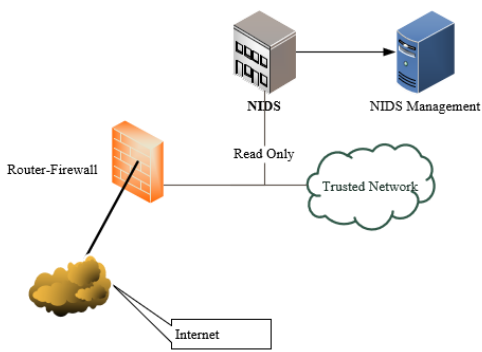

A network - based detection system can be connected to a wide range of network applications like ethernet. These stand - alone devices are connected to the network to monitor network traffic. The NIDS sensors attached to the network will monitor the network traffic from all the devices on the network. For example, the NIDS sensors are installed on the subnet where firewalls are located to detect threats like Denial of Service (DoS) and other such attacks.

### 3) Host - based Intrusion Detection Systems (HIDS)

The key objective of Host - based intrusion detection systems (HIDS) is to collect information about activities on a single system or host.

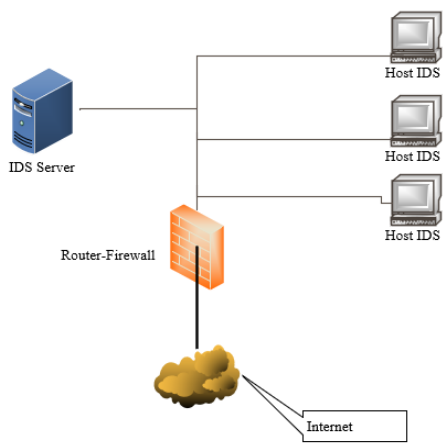

Host Protection is the technique and technologies utilized to protect Hosts that connect to Enterprise networks to protect corporate data and consumer's data security and privacy. Next - Generation host protection systems must proactively identify host systems likely to be impacted by adversarial threats, proactively protect against emanating malicious threats, proactively Detect potential venomous malware via containerization techniques, proactively respond to an ever - growing and mutating threat landscape and proactively take measures to ensure recovery of systems with minimal impact to overall business operations.

### 4) Signature - Based Intrusion Detection Systems

The signature - based IDS analyses each packet's content at Layer 7 and compares it with a set of pre - defined signatures. The signatures of identities need to be regularly updated to ensure that the resource intruders are current. The framework is useful only if the database is up to date to current at a given point in time. The hackers can bypass the signature - based IDS by recurrently modifying minor things about the attach and keeping the database signatures out of place. The approach is effective when detecting attacks that use pre - defined patterns.

### 5) Anomaly - based Intrusion Detection Systems

Anomaly - based detection generally needs to work on a statistically significant number of packets because any packet is only an anomaly compared to some baseline (Timothy, 2014). In an anomaly - based access system, the access system detects both network and computer access and misuses by monitoring system activity and classifying it as usual or undesirable.

### b) Network Intrusion Prevention Steps

### 1) Defining Scope and Priorities

Defining the scope and priorities the process and controls in place ensure that the information assets are secure are critical to an organization. The procedure provides logged items requirements to enable security alerting and behavioral security analytics for detecting attacks and inappropriate actions. The procedure also includes monitoring requirements based on the security logging performed by application developers, system integrators, project teams, and capability owners to enable security alerting and behavioral analytics to detect attacks and inappropriate actions. This includes all information and network assets, including but not limited to critical applications, systems, services, and networks that handle confidential information, accept network connections, or make access control – authentication and authorization decisions must be recorded and retain audit - logging information.

### 2) Institute Monitoring

Once the scope and prioritization are baselined, the required processes and controls need to be implemented are the following.

- Establish security logging requirements based on enterprise, application, and project scope (e. g., classification, criticality)
- Identify items to be logged and configure the IDS frameworks.
- Define criteria for alerts based on logged items.
- Identify the length of time the data must be held.
- Send logs to a central monitoring IDS system that validates the network traffic and identifies any threats or vulnerabilities.
- Validate logs are parsed and indexed correctly in the central monitoring system.
- Alerting for unexpected activity –to include alerting customers of unexpected activity on their accounts
- Validate expected content is what is being captured.
- Validate that alerts are functioning as expected within the environments targeted for security monitoring.
- Setup monitoring rules to trigger and report suspicious events.

### 3) Distributed IDS/IPS Architectures

Hybrid or distributed IDS/IPS framework consolidates the data from different sources or devices connected to the sensors into a secure and centralized repository. The approach combines the properties of host - based and network - based frameworks and complements the signature - based and anomaly - based system by combining the features and improve algorithms by supervised learnings.

Supervised learnings provide more flexibility than traditional signature - based IDS systems by categorizing the know attacks through identifying anomalous activities.

### 4) Fine Tuning and Additional Considerations

The objective of tuning the detection sensors is to reduce the false positives and increase threat alerts' accuracy. However, there are no pre - defined steps to fine - tune the IDS and capture all the threats and vulnerabilities. A few methods that can be followed to improve the efficiency and reliability if IDS is by consistently updating and refining the pre - defined alerts based on data collected from different sources, ensure to have IDS sensors on the right location in the network with essential features to detect and prevent the threats and attacks, etc.

### 5) Training and provide awareness

The employees should be performing different roles in an organization and have different access permission based on the role and the type of work they do. Effective training processes need to be developed to ensure that the users know the enterprise security policies, the resources they need, access to, usage of company - provided hardware assets, etc. The policies need to be updated regularly, and training needs to be conducted to avoid insider threats, social engineering, etc. The security policies need to be documented and made available to add resources and implement a monitoring mechanism to ensure that the policies ad standards are being followed.

### 6) Conclusion

As The Entity Proxy migrates from the broken "Trust but Verify" information security model towards the "Zero Trust" information security model, the Company must ensure that all hosts are accessed securely regardless of location, as well as ensure that host's access control should be based on a "Need to Know" basis and should be strictly enforced. In addition, Entity Proxy needs to move towards inspection and logging of all network traffic on the Company's network for proactive monitoring and logging to ensure the integrity of confidentiality of corporate data and protect its consumer's data privacy and security.

The digital forensics process should have all processes and control in place to alert before a breach occurs. To accomplish this, the Company needs to install IDS sensors at the right places to monitor the network traffic and collect the logs. It is recommended to use a hybrid framework to leverage the features of host - based and network - based IDS/IPS frameworks. Network security procedures need to be baselined. A robust change management process needs to be implemented to revisit the policies at regular intervals and update the monitoring algorithms to reduce the false positives and improve the monitoring systems' efficiency. The incident containment plan must be developed, considering different types of incidents such as malware infection, DDoS, or an attacker with active sessions on machines. System security must be hardened to provide only necessary ports, protocols, and services to meet business needs and support technical controls such as antivirus, file integrity monitoring, and logging as part of the baseline operating build standard. The physical security should be useful to reduce authorized physical access, and all network and information systems used for Entity proxy, in conjunction with the terms of contractual agreements, must be auditable. The enterprise should have processes to perform different forensic investigation phases like identification, Assessment, Remediation, Measures and Measurement, and Communication procedures to protect the information assets and ensure business continuity.

## 5. Appendix A: Risk Mitigation Techniques and Plans

### 1) Risk Mitigation Strategy Development

Risk mitigation is the process of developing actionable insights that reduce threats to the overall well - being of an organization (LeClair, 2019). The management and senior leadership team are primarily responsible for managing the risks associated with different phases of resources in the Company. Effective enterprise risk management is a competitive requirement and an important part of achieving our objectives and protecting and enhancing our corporate reputation. In the context of realizing strategic objectives, the Company realizes that risks should be taken in a deliberate and controlled manner with ownership and accountability.

### 2) The Risk Mitigation Plan

The Enterprise Risk Management (ERM) team is a service and governance function to all areas of the Company, and is responsible for developing, deploying, documenting, and managing scalable and sustainable enterprise risk management plan. This includes the Company's overall risk management program that enables it to understand, manage the risks and create a risk mitigation plan to identify, evaluate and mitigate all the risks at different phases like initiation, planning, and implementation.

### 3) Risk Mitigation Types

Risk response should be considered during the enterprise risk assessment. It is an essential component of the risk assessment process to recommend and request a response to the risks identified. The following are the risk mitigation types.

- Risk Avoidance: Removal of the risk by eliminating the source of risk or avoiding the activity that carries this risk.
- Risk Reduction: Risk depends on the impact (consequences) and likelihood (probability); hence, risk reduction reduces the impact and likelihood or both by implementing compensating controls.
- Risk Transfer: Risk transfer is transferring the possible consequences of a risk to another party. However, risk ownership remains on the business owner.
- Risk Acceptance: Every business activity implies risk. Risk acceptance implies the conscious acceptance of the effects of risk by the Company. Accepted risks should be taken in a deliberate and controlled manner with ownership and accountability.

### 4) Phases & Timelines

The risk mitigation process should be performed at all the critical phases like planning, initiation, implementation, and closeout phases at regular intervals. The **initiation phase** is the initial phase of the project where the requirements are assessed, the risks and unknowns are evaluated and communicated to the stakeholders. The risks and benefits are evaluated at this phase of the project before sign - off to the next phase. Once the project sponsor sign - off the project requirements, the **planning** is conducted where the project is broken into actions, and the risk is identified at each activity that impacts the timelines and business. Once the planning is complete, the implementation phase is where the user stories are created for the body of work to be performed and implemented. The closeout phase ensures that all the risks are assessed and mitigated in accordance with proper migration type and strategy.

5) *Mitigating Common Breaches*

The risk prevention plan in the Entity Proxy includes the strategy to mitigate the common breaches that could happen in an enterprise. The remediation process should be the first path for resolution and starts with the Risk Management team developing a remediation plan with the finding owner to determine an agreed - upon course of action to remediate the finding. The following are the mitigation plan to be implemented at Entity Proxy, such as employee awareness training, resource patching, Data Loss Prevention (DLP) integration, Physical Security, VPNs, active IDS/IPS, and proper policies.

6) *Employee Awareness Training*

The objectives for privacy and security training at Entity Proxy are to educate the workforce on authorized use and handling of company's information assets and information systems, inform users of applicable security and privacy policies and the importance of compliance, and help users understand how to recognize and report problems, incidents or violations, and encourage such reporting.

7) *Active IDS/Intrusion Prevention Systems*

Implementing active intrusion detection and prevention systems can help to reduce risks. Entity proxy's host protection strategy should be based on the Zero Trust model of proactive Virus Prevention (VP) and proactive Behavioral - Based Patterns Recognition (BBPR) techniques for threat detection, analysis, and neutralization via inspection of all network traffic, maintaining host integrity protection techniques with the primary objective of protecting Company's data while protecting user's data security and privacy. Host protection involves protecting anything with an IP address. To ensure the integrity of these hosts, it is critical to proactively Identify, Protect, Detect, Respond, and Recover against malicious activity that impacts the host's operating system. It is also vital to monitor all malicious network traffic flowing to/from network and devices.

8) *Timely and Consistent Patch Schedule*

Entity Proxy should further improve risk mitigation by implementing a secure configuration for desktop software such as disabling scripting in Adobe PDF, remove dated software (old java versions, browser plugin - flash, Silverlight, shockwave, java, etc.) from workstations/ laptops. Also, patch management should be ongoing to at minimum cover browser plugins when security releases are available. Laptops and sensitive servers should also have a hardware root - of - trust to ensure critical software components are not being modified, with full disk encryption for servers with sensitive data.

a) *Data Loss Prevention (DLP) System*

Data loss prevention (DLP) is a set of tools and processes used to ensure that sensitive data is not lost, misused, or accessed by unauthorized users (Groot, 2020). A DLP software needs to be implemented at Entity proxy that identifies and classifies the confidential and critical data and identifies the policy violations defined by the security office in accordance with regulatory compliance laws such as CCPA, GDPR, etc.

b) *Physical Security Considerations*

The entity proxy's physical security policies and standards include implementing physical security systems (PSS), including burglar alarms, surveillance cameras, inventory, and cash safes, and storefront security controls.

c) *Virtual Private Networks*

Entity proxy's employees are recommended to connect to the Company's internet network using Virtual Private Networks (VPN). The VPN should be integrated with active directory services. The following are a few benefits of using a VPN solution in the enterprise.
- Allows employees to connect with the Company's internal network.
- It helps admins track the profile and system configurations to ensure that the device is safe from threats, viruses, and malware.
- VPNs are compatible with different platforms.
- Affordable.
- Data transfer will be secure and protected.

The laptops provided by the Company should be provisioned with a VPN Client software like Cisco AnyConnect. Employees are required to login to VPN Server using Active Directory login. It's also recommended to enable multifactor authentication for VPN access as well.

d) *Role - Based Access Control*

Entity Proxy users are given access based on role - based access controls (RBAC) defined in the security standards policy. Every user account is configured with a set of roles, each of which is associated with a specific set of privileges. The role - based access would ensure that the resources have access only to the required applications based on their role and job responsibilities.

9) *Post Incentive Review Procedures*

The Post Incident Review (PIR) process is fundamentally a process for assessing incident response and recovery efforts. Post Incentive Review Procedures begin once the incident is contained and resolved. The entire incident response activities and measures from identification until closing are reviewed, and retrospect is conducted with all concerned stakeholders and document the learnings, what went well, areas of improvement, etc. which could help the Company prepare and respond when a breach occurs in the future.

# References

[1] Business, A. (n. d.). *Incident Response Team: What are the Roles and Responsibilities?* Retrieved from Incident Response Team: What are the Roles and Responsibilities?: https: //cybersecurity. att. com/resource - center/ebook/insider - guide - to - incident - response/arming - your - incident - response - team

[2] CERT Insider Threat Center. (2011). *Insider Threat and Physical Security of Organizations.* Retrieved from Carnegie Mellon University: Software Engineering Institute: https: //insights. sei. cmu. edu/insider - threat/2011/05/insider - threat - and - physical - security - of - organizations. html

[3] Chew, E., Swanson, M., Stine, K., Bartol, N., Brown, A., & Will, R. (2008). *Performance Measurement Guide for Information Security.* Retrieved from National Institue of Standards and Technology: U. S. Department of Commerce: https: //nvlpubs. nist. gov/nistpubs/Legacy/SP/nistspecialpublication800 - 55r1. pdf

[4] Cichonski, P., Millar, T., Grance, T., & Scarfone, K. (2012). *Computer Security Incident Handling Guide.* Retrieved from National Institute of Standards and Technology: U. S. Department of Commerce: https: //nvlpubs. nist. gov/nistpubs/specialpublications/nist. sp.800 - 61r2. pdf

[5] Cloudflare. (n. d.). *What is a DDoS Attack?* Retrieved from What is a DDoS Attack?: https: //www.cloudflare. com/learning/ddos/what - is - a - ddos - attack/

[6] DARPA. (2020). *About Us.* Retrieved from Defense Advanced Research Projects Agency: https: //www.darpa. mil/about - us/about - darpa

[7] Eric Conrad, J. F. (2017). *Network Based Intrusion Detection System.* Retrieved from Network Based Intrusion Detection System: https: //www.sciencedirect. com/topics/computer - science/network - based - intrusion - detection - system

[8] Gladyshev, P. (2004). *Formalising Event Reconstruction in Digital Investigations.* Retrieved from Research Gate: https: //www.researchgate. net/profile/Pavel_Gladyshev/publication/335444758_Formalising_Event_Reconstruction_in_Digital_Investigations/links/5d665d4e92851c70c4c3a48f/Formalising - Event - Reconstruction - in - Digital - Investigations. pdf

[9] Glass - Vanderlan, T. R., Iannacone, M. D., Vincent, M. S., Chen, Q. G., & Bridges, R. A. (2018). *A Survey of Intrusion Detection Systems Leveraging Host Data.* Retrieved from Cornell University: https: //arxiv. org/pdf/1805.06070. pdf

[10] Gordon, W. J., Wright, A., Aiyagari, R., Corbo, L., Glynn, R. J., Kadakia, J.,. . . Landman, A. B. (2019). *Assessment of Employee Susceptibility to Phishing Attacks at US Health Care Institutions.* Retrieved from National Center for Biotechnology Information: https: //www.ncbi. nlm. nih. gov/pmc/articles/PMC6484661/

[11] Groot, J. D. (2020). *What is Data Loss Prevention (DLP) ? A Definition of Data Loss Prevention.* Retrieved from What is Data Loss Prevention (DLP) ? A Definition of Data Loss Prevention: https: //digitalguardian. com/blog/what - data - loss - prevention - dlp - definition - data - loss - prevention#:~: text=Data%20loss%20prevention%20 (DLP) %20is, or%20accessed%20by%20unauthorized%20users. &text=DLP%20also%20provides%20reporting%20to, for%20forensics%20and%20inc

[12] Hayes, D. R. (2015). *The Scope of Computer Forensics.* Retrieved from A Practical Guide to Computer Forensics Investigations: https: //coloradotech. vitalsource. com/#/books/9780132756150?context_token=caad0870 - f0bf - 0138 - 6669 - 52415a6fb4a2

[13] HOSPELHORN, S. (2020). *What is The Cyber Kill Chain and How to Use it Effectively.* Retrieved from What is The Cyber Kill Chain and How to Use it Effectively: https: //www.varonis. com/blog/cyber - kill - chain/

[14] Kral, P. (2011). *Incident Handler's Handbook.* Retrieved from SANS Institute Information Security Reading Room: https: //www.sans. org/reading - room/whitepapers/incident/incident - handlers - handbook - 33901

[15] LeClair, L. (2019). *How to Develop a Mitigation Strategy.* Retrieved from How to Develop a Mitigation Strategy: https: //blog. newcloudnetworks. com/how - to - develop - a - mitigation - strategy

[16] Mashechkin, L., Petrovskiy, M., Popov, D., & Tsarev, D. (2015). Applying Text Mining Methods for Data Loss Prevention. *Programming and Computer Software, 41* (1), 23 - 30.

[17] MUSE. (n. d.). *Digital Forensics Overview.* Retrieved from Digital Forensics Overview: https: //class. ctuonline. edu/_layouts/MUSEViewer/MUSE. aspx?mid=18546211

[18] MUSE. (n. d.). *Network and Internet Breaches.* Retrieved from MUSE: https: //class. ctuonline. edu/_layouts/MUSEViewer/MUSE. aspx?mid=18546342

[19] MUSE. (n. d.). *Threats and Vulnerabilities.* Retrieved from MUSE: https: //class. ctuonline. edu/_layouts/MUSEViewer/MUSE. aspx?mid=18546253

[20] Spitzner, L. (2019). *Applying Security Awareness to the Cyber Kill Chain.* Retrieved from SANS: Cyber Kill Chain Security Awareness Planning: https: //www.sans. org/security - awareness - training/blog/applying - security - awareness - cyber - kill - chain

[21] Stergiopoulos, G., Kotzanikolaou, P., Theocharidou, M., & Gritzalis, D. (2015). Risk mitigation strategies for critical infrastructures based on graph centrality analysis. *International Journal of Critical Infrastructure Protection, 10*, 34 - 44.

[22] Timothy J. Shimeall, J. M. (2014). *Anomaly - Based Detection.* Retrieved from Anomaly - Based Detection: https: //www.sciencedirect. com/topics/computer - science/anomaly - based - detection

[23] Tomur, E., & Erten, Y. (2006). Application of Temporal and Spatial Role Based Access Control in 802.11 Wireless Networks. *Computers & Security, 25* (6), 452 - 458.

[24] Watt, A. (2019). *Project Management.* Creative Commons Attribution.

[25] Yeo, L. H., Che, X., & Lakkaraju, S. (2017). *Understanding Modern Intrusion Detection Systems: A Survey.* Retrieved from Cornell University: https: //arxiv. org/pdf/1708.07174. pdf